\documentclass[11pt]{article}

\usepackage[final]{acl}
\usepackage{times}
\usepackage{latexsym}

\usepackage[T1]{fontenc}

\usepackage[utf8]{inputenc}

\usepackage{microtype}
\usepackage{adjustbox}
\usepackage{booktabs}
\usepackage{multirow}
\usepackage{subcaption}
\usepackage{amsmath}
\usepackage{tcolorbox}
\usepackage[table]{xcolor}

\usepackage{array} 

\usepackage[hypcap=false]{caption} 

\usepackage{natbib}
\tcbuselibrary{listings}

\usepackage{inconsolata}

\usepackage{graphicx}
\usepackage{enumitem}

\title{SWE-Tester: Training Open-Source LLMs for Issue Reproduction in Real-World Repositories}

\author{
  \textbf{Aditya Bharat Soni\textsuperscript{1}} \thanks{\ \ Work done during an internship at Nutanix.},
  \textbf{Rajat Ghosh\textsuperscript{2}},
  \textbf{Vaishnavi Bhargava\textsuperscript{2}},
  \textbf{Valerie Chen\textsuperscript{3}},
  \textbf{Debojyoti Dutta\textsuperscript{2}}
\\
  \textsuperscript{1}Carnegie Mellon University,
  \textsuperscript{2}Nutanix,
  \textsuperscript{3}Stanford University
\\
  \small{
    \textbf{Correspondence:}
    \href{mailto:adityabs@cs.cmu.edu}{adityabs@cs.cmu.edu},
    \href{mailto:rajat.ghosh@nutanix.com}{rajat.ghosh@nutanix.com},
    \href{mailto:vaishnavi.bhargava@nutanix.com}{vaishnavi.bhargava@nutanix.com}
  }
\\
  \small{
  }
}

\begin{document}
\maketitle
\begin{abstract}
Software testing is essential for maintaining software reliability, yet issue reproduction tests are often missing when bugs are reported. Automatically generating such tests can accelerate debugging, support test-driven development, and improve automated software engineering systems that rely on execution-based verification. Existing approaches largely depend on proprietary language models, limiting accessibility and reproducibility. We present \textbf{SWE-Tester} \footnote{https://github.com/adityasoni9998/SWE-Tester} a framework for training open-source large language models to generate issue reproduction tests from natural language issue reports. To support this task, we curate a dataset of \textbf{41K} training instances from \textbf{2.6K} open-source GitHub repositories. Fine-tuning on this dataset improves issue reproduction performance across multiple model families, yielding gains of up to \textbf{10\%} points in success rate and \textbf{21\%} in change coverage on SWT-Bench Verified. We further study the effects of model size, training data scale, and inference-time compute, observing consistent improvements along all three axes. Our results demonstrate that open-source models can be effectively specialized for issue reproduction test generation and provide a scalable foundation for future software testing and automated issue-resolution systems.

\end{abstract}

\section{Introduction}\label{sec:introduction}
Large language models have demonstrated remarkable improvements on software engineering (SWE) tasks, as evidenced by progress on benchmarks like SWE-Bench \citep{jimenez2024swebench}, which evaluates an LLM's ability to resolve codebase issues. However, issue resolution is only one component of the software development lifecycle. Rigorous software testing is equally critical, yet developers often find writing reproduction tests tedious \citep{straubinger2023survey}, and such tests are rarely present when a bug is reported \citep{kang2023large,mundler2024swtbench}. 

Automating the generation of bug-reproducing tests can significantly enhance developer productivity, enable test-driven development (TDD) \citep{beck2022test}, and boost automated bug-fixing pipelines by serving as localized execution-based verifiers \citep{xia2024agentless,jain2025regym,wei2025swerl}. Despite these benefits, existing state-of-the-art bug reproduction techniques, such as AssertFlip \citep{khatib2025assertflip} and e-Otter++, rely heavily on expensive, closed-source APIs (e.g., GPT-4o, Claude-3.7-Sonnet). In industrial environments, sending proprietary code to third-party APIs poses severe intellectual property and data privacy risks. Consequently, training local, open-source models for repository-level test generation is a critical operational necessity, yet it remains largely unexplored.

Furthermore, repository-level test reproduction is distinct from source-code editing. \citet{mundler2024swtbench} observe that there is no correlation between an LLM's ability to fix a bug and its ability to reproduce it, making test reproduction a unique and challenging task that requires targeted training. 

\begin{figure}[htbp]
    \centering
    \includegraphics[height=3.5cm,width=0.48\textwidth]{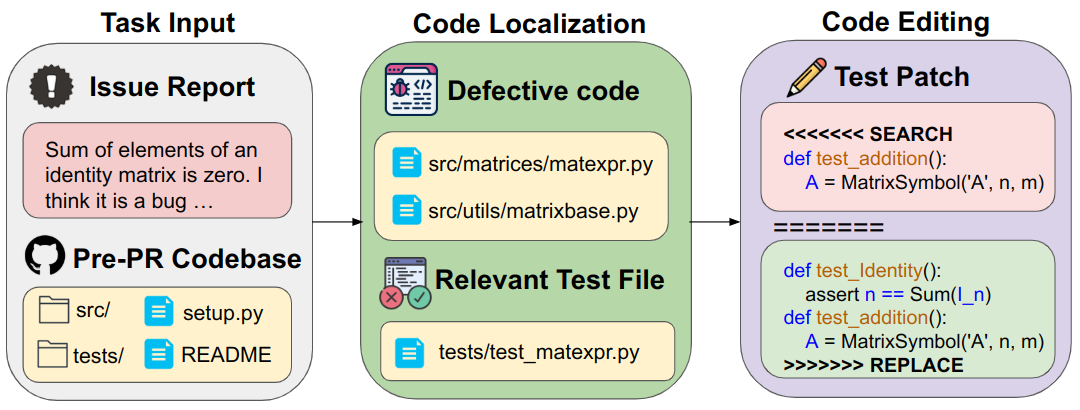}
    \caption{\textbf{Our proposed SWE-Tester pipeline.} Given an issue description and the pre-PR repository (which has this issue/bug), we first retrieve defective source code and relevant test file (code localization). Next, we edit the test file to augment it with reproduction tests (code editing).}
    \label{fig:system_diagram}
\end{figure}

To address these challenges, we propose \textbf{SWE-Tester} -- a localized, privacy-preserving framework for generating issue reproduction tests using fully open-source, open-weights LLMs. As shown in Figure \ref{fig:system_diagram}, our framework employs a standardized two-step pipeline consisting of code localization followed by code editing. Our analysis reveals that while open-weights models localize code reasonably well, they heavily struggle with editing test files. Since editing is the primary bottleneck, we focus on fine-tuning open-source LLMs to improve their repository-level editing performance. Our contributions are summarized as follows:
\begin{enumerate}[leftmargin=*]
  \item \textbf{Privacy-Preserving Training Dataset}: We leverage raw, open-source GitHub pull requests to curate a high-quality dataset of 41K training instances from 2.6K repositories, allowing local training without exposing proprietary code.
  \item \textbf{Broad Generalization Across Open Model Families}: We fine-tune five open-source LLMs across varying parameter sizes and architectures: Qwen-2.5-Coder (7B, 14B, and 32B), Llama-3.1-8B, and Gemma-3-12B. Fine-tuning yields absolute gains of up to 10\% in success rate and 21\% in change coverage on SWT-Bench Verified, outperforming strong closed-source baselines like ZeroShotPlus \citep{mundler2024swtbench}.
  \item \textbf{Operational Cost and Scaling Analysis}: We analyze the trade-offs of inference-time compute scaling (sampling up to 32 candidate patches). Our results show consistent performance gains without saturation, offering concrete guidance on balancing execution latency, token costs, and safety sandboxing in enterprise CI/CD pipelines.
\end{enumerate}
\section{Related Work}\label{sec:related_work}

\textbf{Issue Reproduction Benchmarks and Training.}
Issue reproduction is substantially more challenging than conventional unit test generation because tests must be derived from a natural language issue description and a buggy repository without access to the issue-fixing patch. SWT-Bench \citep{mundler2024swtbench} and Defects4J \citep{just2014defects4j} are widely used benchmarks for evaluating this capability. While SWT-Bench supports both \textit{unit test} and \textit{reproduction script} settings, prior training efforts primarily focus on reproduction scripts or depend heavily on proprietary models. For example, \citet{jain2025regym} train testing agents using trajectories generated by closed-source LLMs, while \citet{SWESwiss2025} employ reinforcement learning on pipeline-based workflows. In contrast, we train open-source LLMs directly for the more realistic \textit{unit test} setting using developer-written tests mined from historical pull requests.

\textbf{Reproduction Test Generation Workflows.}
Existing approaches typically employ multi-stage pipelines or agentic workflows involving planning, localization, root-cause analysis, and iterative refinement \citep{ahmed2025otter,kang2023large,wang2024aegisagentbasedframeworkgeneral,mundler2024swtbench,khatib2025assertflip}. Although effective, these systems frequently rely on proprietary LLMs and generate intermediate reasoning steps for which supervised training data is unavailable. We instead adopt a simple two-stage workflow consisting of localization and editing, enabling straightforward supervised fine-tuning from raw pull request data.

\textbf{Automated Issue Resolution.}
Issue resolution focuses on modifying source code to fix bugs and is commonly evaluated on SWE-Bench and its variants \citep{jimenez2024swebench,yang2025swebench}. Recent training approaches either train LLM agents \citep{pan2025training,jain2025regym,yang2025swesmith} or specialized pipeline components \citep{wei2025swerl,xie2025swefixer,SWESwiss2025}. We leverage the large-scale pull request data collected by these efforts but repurpose it to train models for issue reproduction, a related yet distinct task.

\section{Methodology}
In this section, we formally define the issue reproduction task in \S\ref{sec:task_definition}, describe our workflow in \S\ref{sec:workflow}, explain our data curation and training methodology in \S\ref{sec:train_data_curation}, and finally describe the inference scaffold used to leverage inference-time compute in \S\ref{sec:inference_scaffolds}.

\subsection{Issue Reproduction Task}\label{sec:task_definition}
We formally define the issue reproduction task following  \citet{mundler2024swtbench}. Given a code repository $\mathcal{R}$, a natural language issue description $\mathcal{I}$ , which is a bug report or a feature request in $\mathcal{R}$, and a golden code patch $\mathcal{X^{*}}$, containing source code changes from the issue-resolving PR. We denote the codebase $\mathcal{R}$ after applying a patch $\mathcal{X}$ as $\mathcal{R}$ $o$ $\mathcal{X}$. A test $s$ can either pass (\textit{P}) or fail (\textit{F)} when executed on $\mathcal{R}$. A test $s$ fails if its execution triggers an error (for eg: \texttt{AssertionError}, \texttt{ValueError}, etc.), and it passes otherwise. A test $s$ reproduces an issue $\mathcal{I}$ if it fails on the pre-PR codebase $\mathcal{R}$ and passes on the post-PR codebase $\mathcal{R}$ $o$ $\mathcal{X^{*}}$. Such tests are defined as fail-to-pass tests ($\textit{F} \rightarrow \textit{P}$), and $\textit{F} \rightarrow \textit{F}$, $\textit{P} \rightarrow \textit{P}$ and $\textit{P} \rightarrow \textit{F}$ tests are defined similarly. A set of tests \textit{T} reproduces an issue if: at least one test in \textit{T} is $\textit{F} \rightarrow \textit{P}$ (i.e it correctly exposes the issue), and no test in \textit{T} fails on $\mathcal{R}$ $o$ $\mathcal{X^{*}}$ (avoiding $\textit{F} \rightarrow \textit{F}$
and $\textit{P} \rightarrow \textit{F}$ tests with incorrect testing logic or syntax issues). Also, this allows $\textit{P} \rightarrow \textit{P}$ regression tests in \textit{T}. Given the pre-PR codebase $\mathcal{R}$ and the issue description $\mathcal{I}$, the task of the LLM is to generate a test set \textit{T} that reproduces the issue. 

\subsection{Workflow for Issue Reproduction}\label{sec:workflow}
In this section, we describe our two-step workflow for generating issue reproduction tests (Figure \ref{fig:system_diagram}) using \textbf{fully open-source LLMs}. The first step is code localization -- retrieving relevant source code and test code from the pre-PR codebase. The second step is code editing -- generating a patch to modify the retrieved test code and augmenting it with issue reproduction tests. 
\subsubsection{Code Localization}\label{sec:code_localization}
For code localization, we retrieve potentially defective source code files and a relevant test file from the pre-PR codebase. The source code files give context on possible root causes and highlight segments that require further testing for issue reproduction. The test file helps the LLM to understand existing testing conventions, identify gaps in testing logic critical for issue reproduction, and reuse relevant utility functions and classes.

Prior pipeline-based approaches follow similar localization steps, with some localizing to more fine-grained locations (e.g., classes or functions) by prompting an LLM \citep{khatib2025assertflip,ahmed2025otter}. We provide the complete content of the retrieved files following training methods for issue resolution, which find that including the entire file contents improves performance by teaching the model to identify relevant code segments \citep{xie2025swefixer,wei2025swerl}. For test code, we follow prior work \citep{ahmed2024tdd,ahmed2025otter,ahmed2025executionfeedbackdriventestgeneration}, that achieve strong performance on this task, and localize a single test file. While this may miss scenarios like utility code and tests being edited in separate files, it simplifies localization during inference and aids training data filtering (\S\ref{sec:train_data_curation}).

For source code, we adopt the coarse-to-fine retrieval strategy of \citet{xie2025swefixer}: BM25 \citep{robertson2009probabilistic} retrieves the top-30 relevant source code files using issue description as the query, and a fine-tuned Qwen2.5-7B \citep{qwen2.5} model then selects the defective files given skeletons of these 30 files. For test code, we use BM25 to retrieve the top-30 candidate test files with issue being the query, which are then reranked using Qwen3-Embedding-8B \citep{zhang2025qwen3} model. Since the top-ranked file is not always correct, we use an inference scaffold (\S\ref{sec:inference_scaffolds}) that samples multiple test patches from the LLM for the top-K test files and reranks them to select the most appropriate patch. Although we can train a retriever model for test files, we find code editing step to be the main bottleneck (\S\ref{sec:retrieval_results}).
\subsubsection{Code Editing}\label{sec:code_editing}
The next step is code editing, where the LLM generates a patch to add reproduction tests to the test file given the issue and the localized context. A key design choice is the patch format: while the unified diff format (used in \texttt{git diff}) is a standard representation, it is highly sensitive to mis-specifications since LLMs often struggle with precise line numbers \citep{jimenez2024swebench, mundler2024swtbench}. \citet{mundler2024swtbench,ahmed2025otter} use LLM-friendly formats that only allow insertion and replacement of functions, but they do not support editing other code segments like imports,
global constants, class attributes, etc. This is particularly problematic for training data curation because edits to the test suite in developer-written PRs are not restricted to function-level edits. To overcome these limitations, we use the Search/Replace format \citep{xia2024agentless,wei2025swerl}, which can represent any arbitrary file edits (example shown in Appendix \ref{app:prompt_issue_reproduction}) and simplifies training data curation \S\ref{sec:train_data_curation}.
\subsection{Dataset Curation and Training Methodology}\label{sec:train_data_curation}
Next, we describe our data curation and training methodology for the code editing step. Prior training methods for issue resolution have scraped raw PR data from GitHub repositories, with some also developing sandboxed execution environments \citep{pan2025training,jain2025regym}. While they target issue resolution, we instead train models to predict the test patch for issue reproduction -- a related but distinct task, since an LLM’s ability to fix an issue does not correlate with its ability to reproduce it (\S\ref{sec:introduction}). During training, the LLM is prompted with the issue description, the pre-PR contents of the gold defective source code and gold test files edited by the issue-resolving PR. Refer to Appendix \ref{app:prompt_issue_reproduction} for the exact prompts used in our work.

To curate training data, we leverage PR data curated by SWE-Gym \citep{pan2025training}, SWE-Fixer \citep{xie2025swefixer}, SWE-rebench \citep{badertdinov2025swerebenchautomatedpipelinetask}, and SWE-Bench Extra \citep{badertdinov2024scaling}, which contain a total of 140K issue–PR pairs across 5K repositories. However, these datasets do not provide pre-PR file contents of the source code and test files edited by the PR, so we scrape them by cloning the pre-PR codebase for each instance. Since raw PR data is very noisy, we exclude PRs that satisfy any of these conditions: 1) belongs to repositories present in SWT-Bench (our evaluation set), 2) edits non-Python files, 3) edits more than three or zero source code files, 4) does not edit \textit{exactly} one test file, or 5) has empty issue descriptions. Since not every edit made by the test patch adds a reproduction test, filtering criterion 4) helps ensure that the edits are \textit{generally} relevant to the issue. We further discuss our rationale behind filtering criterion in Appendix \ref{app:filtering_rationale}. After filtering and de-duplication, we obtain \textbf{41K} high-quality training instances from \textbf{2.6K} repositories. We write custom scripts to translate developer-written test patches from the unified diff format to Search/Replace format. Table \ref{table:dataset-stats-combined} shows dataset statistics and repo-level instance frequencies. Similar to our parent datasets, our data also has a long-tail distribution: while \textbf{70}\% repositories have $<$5 instances, the most frequent repository (\texttt{pandas}) has 1529 instances.

\subsection{Inference Scaffold}\label{sec:inference_scaffolds}
Our localization pipeline (\S\ref{sec:code_localization}) retrieves defective source files using SWE-Fixer 7B Retriever \cite{xie2025swefixer} and reranks test files with Qwen3-Embedding-8B \citep{zhang2025qwen3}. Since the top-ranked test file may not be correct, we generate \textit{N} = 8 patches with temperature 0.5 for each of the top-\textit{K} = 4 test files, yielding 32 candidates. All candidates use the same source code context from SWE-Fixer, as prior work \citep{mundler2024swtbench,ahmed2025otter} shows test localization has greater impact on model performance than source localization.

From these 32 candidates, we select the most appropriate patch through a multi-step filtering and reranking process. First, we discard all patches that are not applicable to the test file. Next, for each applicable patch, we execute the original test file and the modified test file (after applying the patch) on the pre-PR codebase. We compare the test execution logs of the two runs and discard all patches which introduce errors (e.g. \texttt{SyntaxError}, \texttt{ImportError}, etc.).

Since a reproduction test set must have at least one test that fails on the pre-PR codebase (\S\ref{sec:task_definition}), we only consider all such patches for the final patch set $\mathcal{S}$. We rerank patches in $\mathcal{S}$ using a self-consistency mechanism similar to \citep{SWESwiss2025}. We assign a score to each patch $p$ in $\mathcal{S}$, computed as:
$\text{Score}(p) = \text{Score}_\text{EM}(p) + \text{Score}_\text{Sim}(p)$. $\text{Score}\text{EM}(p)$ is the standard self-consistency measure, counting how often $p$ occurs in $\mathcal{S}$. $\text{Score}\text{Sim}(p)$ rewards candidates that lie in dense clusters of plausible variants: we compute pairwise similarity between patches with Python’s \texttt{difflib.SequenceMatcher}, identify the top-$r$ nearest neighbors for each candidate, and average their similarity scores. Following \citet{SWESwiss2025}, we set $r=|\mathcal{S}|/2$. The patch with the highest combined score is selected as the final answer. We describe potentially better reranking methods in Appendix \ref{app:reranking_options}

\section{Experimental Setup}\label{sec:experimental_setup}
We perform supervised fine-tuning of Qwen2.5-Coder Instruct (7B, 14B, 32B) \citep{hui2024qwen2}, Llama-3.1-8B Instruct \citep{dubey2024llama}, and Gemma-3-12B Instruct \citep{gemmateam2025gemma3technicalreport}. To fit hardware constraints and preserve high training efficiency, we filter out instances exceeding a 16K token limit (using the Qwen2.5-Coder-7B tokenizer), resulting in 23K active training instances. We train all models for 2 epochs with a batch size of 32, a cosine learning rate scheduler, and the AdamW optimizer. The learning rate is set to $1e^{-5}$ for Gemma-3-12B and $3e^{-5}$ for all other models. Training is performed on 2 to 8 NVIDIA A100 GPUs using the \texttt{torchtune} or \texttt{Llama-Factory} libraries.

We evaluate performance on the SWT-Bench Verified dataset \citep{mundler2024swtbench} (433 instances) using three core metrics: Applicability ($\mathcal{W}$), Success Rate ($\mathcal{S}$), and Change Coverage ($\Delta\mathcal{C}$). Applicability measures the percentage of instances where generated patches apply cleanly. Success Rate is the percentage of instances where the generated tests reproduce the bug ($F \rightarrow P$ on the codebase). Change Coverage represents the fraction of gold patch lines executed by the new tests.
\begin{table}[htbp]
\centering
\caption{Comparing applicability ($\mathcal{W}$), change coverage ($\Delta\mathcal{C}$), and success rate ($\mathcal{S}$) of base and fine-tuned LLMs on SWT-Bench Verified \citep{mundler2024swtbench} using our inference scaffold (\S\ref{sec:inference_scaffolds}).}
\label{tab:model_performance}
\vspace{0.3em}
\renewcommand{\arraystretch}{1.2}
\resizebox{\linewidth}{!}{%
\begin{tabular}{l|ccc|ccc|ccc}
\hline
\hline
\multirow{3}{*}{\textbf{Model}} & \multicolumn{3}{c|}{\textbf{Applicability (\%, $\uparrow$)}} & \multicolumn{3}{c|}{\textbf{Change Coverage (\%, $\uparrow$)}} & \multicolumn{3}{c}{\textbf{Success Rate (\%, $\uparrow$)}} \\
\cline{2-10}
& base & SFT & \textbf{$\Delta$} & base & SFT & \textbf{$\Delta$} & base & SFT & \textbf{$\Delta$} \\
\hline
Llama-3.1-8B-Instruct & 18.01 & 84.76 & \textcolor{green!55!black}{\textbf{+66.75}} & 1.42 & 22.28 & \textcolor{green!55!black}{+20.86} & 0.23 & 6.24 & \textcolor{green!55!black}{+6.01} \\
Gemma-3-12B-Instruct & \textbf{85.91} & \textbf{90.99} & \textcolor{green!55!black}{+5.08} & 5.95 & \textbf{24.12} & \textcolor{green!55!black}{+18.17} & 0.92 & 8.08 & \textcolor{green!55!black}{+7.16} \\
Qwen2.5-Coder-7B-Instruct & 40.42 & 84.99 & \textcolor{green!55!black}{+44.57} & 3.23 & 21.30 & \textcolor{green!55!black}{+18.07} & 1.15 & 8.31 & \textcolor{green!55!black}{+7.16} \\
Qwen2.5-Coder-14B-Instruct & 30.95 & 84.99 & \textcolor{green!55!black}{+54.04} & 2.51 & 23.54 & \textcolor{green!55!black}{\textbf{+21.03}} & 1.39 & 11.79 & \textcolor{green!55!black}{\textbf{+10.40}} \\
Qwen2.5-Coder-32B-Instruct & 84.76 & 84.76 & 0.00 & \textbf{19.14} & 22.73 & \textcolor{green!55!black}{+3.59} & \textbf{9.93} & \textbf{14.09} & \textcolor{green!55!black}{+4.16} \\
\hline
\end{tabular}
}
\end{table}

\section{Results}\label{sec:results}
Table \ref{tab:model_performance} presents our main evaluation results. We outline the key engineering takeaways below:

\textbf{Local SFT consistently improves bug reproduction tests:} Fine-tuning on our curated dataset significantly enhances success rates and change coverage across all five open LLMs. Smaller base models ($\le$14B) struggle with base test generation, achieving $\le$1.39\% success rate. However, SFT yields massive improvements of up to +21.03\% in change coverage and +10.40\% in success rate. While Qwen2.5-Coder-32B shows more modest relative SFT gains (+4.16\% success rate), it achieves the highest overall success rate of 14.09\%. Crucially, these results are achieved using fully open-weights models, offering a privacy-preserving and cost-effective local deployment alternative to closed APIs.

\textbf{Model scale and architecture govern downstream quality:} Fine-tuned performance generally scales with parameter size. Within the Qwen2.5-Coder family, success rates scale predictably from 7B (8.31\%) to 32B (14.09\%). However, base model choice plays an equally vital role; despite their comparable scales, Llama-3.1-8B-Instruct underperforms Qwen2.5-Coder-7B-Instruct by 2.07\% in SFT success rate, and Gemma-3-12B-Instruct underperforms Qwen2.5-Coder-14B-Instruct by 3.71\%.

\textbf{Fine-tuning resolves syntactic patch formatting issues:} A key barrier to automated test execution is generating syntactically valid patches. Base models (excluding 32B and Gemma-3) fail to generate cleanly applicable patches. SFT resolves this formatting bottleneck, boosting applicability to $\ge$84.76\% across all fine-tuned variants. Interestingly, Gemma-3 achieves high applicability (90.99\% SFT) but a lower success rate (8.08\%), highlighting that generating well-formed patches is a necessary but insufficient condition for successful bug reproduction.

While the focus of our work is to investigate how we can improve the performance of open-source LLMs for this task, we include comparisons with prior methods in Appendix \ref{app:prior_methods}.

\subsection{Retrieval Performance for Code Localization}\label{sec:retrieval_results}
These results evaluate our retrieval methods and demonstrate that code editing (rather than localization) remains the primary bottleneck for bug reproduction.

\textbf{Source Code Localization} \quad Table~\ref{tab:retrieval} evaluates source file localization on SWT-Bench Verified. The fine-tuned SWE-Fixer retriever \citep{xie2025swefixer} achieves high precision (58.7\%) and recall (58.7\%). While BM25 Top-30 achieves a higher recall (87.7\%), its extremely low precision (3.4\%) introduces excessive context noise, justifying our coarse-to-fine localization strategy.

\begin{table}[htbp]
\centering
\caption{Source code localization performance on SWT-Bench Verified.}
\label{tab:retrieval}
\small
\setlength{\tabcolsep}{8pt}
\begin{tabular}{lcc}
\toprule
\textbf{Method} & \textbf{Precision (\%)} & \textbf{Recall (\%)} \\
\midrule
BM25 Top-1 & 39.6 & 43.6 \\
BM25 Top-4 & 16.9 & 59.7 \\
BM25 Top-30 & 3.4 & \textbf{87.7} \\
SWE-Fixer Retrieval & \textbf{58.7} & 58.7 \\
\bottomrule
\end{tabular}
\end{table}

\textbf{Test Localization} \quad Table \ref{tab:test_localization} evaluates single test file retrieval. Reranking BM25 candidates with Qwen3-Embedding \citep{zhang2025qwen3} substantially increases recall for small $K$. Reranking with the 0.6B model achieves up to 73.0\% recall, performing almost identically to the 8B model while significantly lowering inference latency and deployment costs.

\begin{table}[htbp]
\centering
\caption{Test file localization comparing BM25 with BM25 Top-30 + Qwen3-Embedding reranking (all metrics in \%).}
\label{tab:test_localization}
\vspace{0.3em}
\resizebox{\columnwidth}{!}{%
\begin{tabular}{@{}c|cc|cc|cc@{}}
\toprule
\multicolumn{1}{c|}{\multirow{3}{*}{Top-$K$}} & \multicolumn{2}{c|}{BM25} & \multicolumn{4}{c}{BM25 Top-30 + Reranking} \\
\cmidrule(lr){2-3} \cmidrule(lr){4-7}
 &  &  & \multicolumn{2}{c|}{0.6B} & \multicolumn{2}{c}{8B} \\
\cmidrule(r){4-5} \cmidrule(l){6-7}
 &   Prec. &   Rec.   & Prec. & Rec. & Prec. & Rec. \\
\midrule
1   & 30.0 & 27.2 & 51.5 & 46.6 & 53.3 & 47.9 \\
2   & 22.7 & 39.7 & 32.6 & 57.4 & 35.0 & 60.7 \\
4   & 15.6 & 53.5 & 19.2 & 65.4 & 19.7 & 67.0 \\
8   &  9.8 & 64.7 & 10.9 & 70.9 & 10.9 & 71.9 \\
16  &  5.9 & 74.1 & 5.9 & 73.0 & 5.9 & 73.1 \\
\bottomrule
\end{tabular}
}
\end{table}
\section{Analysis}\label{sec:analysis}
In this section, we analyze the operational factors that influence the performance and practical deployment of open-weights models for bug reproduction. We evaluate the impact of inference-time compute scaling in \S\ref{sec:inference_scaling} and analyze repository-level performance variations in \S\ref{sec:repo_analysis}. Additional evaluations of offline training data scaling, parameter-efficient fine-tuning (LoRA), and edit-only "oracle" baseline comparisons are detailed in Appendices \ref{app:oracle_results}, \ref{app:train_scaling}, and \ref{app:peft_eval}.

\subsection{Effect of Scaling Inference-Time Compute}\label{sec:inference_scaling}
We evaluate how model performance scales with increased test-time compute. We hold the number of retrieved test files constant ($K = 4$) and vary the number of sampled patches per file ($N \in \{1, 2, 4, 8\}$), resulting in a budget of 4, 8, 16, and 32 candidate patches per issue. The final reproduction patch is selected using our execution-backed filtering and self-consistency scaffold (\S\ref{sec:inference_scaffolds}).

As shown in Figure \ref{fig:inference_scaling}, we observe consistent, monotonic performance gains across the Qwen2.5-Coder family (7B, 14B, and 32B). Notably, neither success rate ($\mathcal{S}$) nor change coverage ($\Delta\mathcal{C}$) shows signs of saturation up to 32 candidate samples. This trend demonstrates that test-time compute scaling is a highly viable strategy for local enterprise deployments: organizations can directly trade off additional local GPU compute cycles for significantly higher test generation quality, all while maintaining complete data privacy.

\subsection{Analyzing Performance Across Repositories}\label{sec:repo_analysis}
Using the inference setup from \S\ref{sec:results}, we analyze the distribution of successfully reproduced bugs across different software repositories to identify domain-specific model behavior. Detailed repository-level breakdown metrics are presented in Table \ref{tab:model-repo-percentages} in Appendix \ref{app:repo_breakdown}.

We find that repository characteristics heavily influence model performance. All fine-tuned models demonstrate high proficiency on lightweight web frameworks such as \texttt{flask} (achieving up to 100\% success rate) but struggle on complex, highly integrated documentation engines like \texttt{sphinx} (under 4\% success rate) and plotting libraries like \texttt{seaborn} (0\% success rate). This performance gap is primarily driven by: (1) environment and dependency complexity, where graphics-rendering or specialized CLI dependencies are harder to initialize in sandbox environments, and (2) long-context dependencies, where complex systems require larger context horizons to localize relevant helper functions. Interestingly, model scaling is not strictly monotonic; the 7B model outperforms the larger 14B and 32B variants on data-focused repositories such as \texttt{xarray} (26.67\% vs. 6.67\%), indicating that localized file conventions are occasionally captured better by smaller, highly focused SFT checkpoints.
\section{Limitations}
While our work demonstrates strong results for training LLMs for issue reproduction, it has several limitations.

First, we evaluate only Python repositories due to the lack of large-scale open-source benchmarks for other languages (e.g., only Defects4J \citep{just2014defects4j} for Java and SWT-Bench \citep{mundler2024swtbench} for Python). 

Second, our training data is constructed using heuristic filtering under computational constraints, and we do not extensively study alternative filtering strategies. We release our data curation pipeline to support future exploration.

Third, we do not explicitly train models for test localization, instead relying on off-the-shelf LLMs, and similarly do not optimize our inference scaffold for best-in-class reranking or patch filtering, as our focus is on studying scalable training for issue reproduction rather than system optimization.

Fourth, we do not verify whether all training patches truly reproduce the underlying issue due to the lack of fully sandboxed execution environments in a large portion of the dataset (notably SWE-Fixer-derived data \citep{xie2025swefixer}). Building such environments is expensive or requires proprietary agent systems.

Finally, we assume single-test-file edits, which simplifies the task but limits applicability to multi-file test refactoring scenarios. Extending to multi-file settings and improving performance via stronger inference-time reasoning or reinforcement learning remains future work.

\bibliography{custom}

\appendix
\begin{table*}[!th]
\begin{minipage}{0.45\textwidth}
\centering
\begin{adjustbox}{max width=\textwidth}
\footnotesize
\begin{tabular}{l l r}
\toprule
\textbf{Category} & \textbf{Metric} & \textbf{Dataset} \\
\midrule
Size & \# Instances & 41K \\
 & \# Repos & 2.6K \\
\midrule
Issue Text & \# words & 175.7 \\
\midrule
\multirow{2}{*}{Gold code patch} & \# Files edited & 1.4 \\
                                       & \# Lines edited & 37.2 \\
\midrule
\multirow{2}{*}{Gold test patch} & \# Files edited & 1 \\
                                & \# Lines edited & 47.9 \\
\bottomrule
\end{tabular}
\end{adjustbox}
\label{table:dataset-stats}
\end{minipage} \hfill \begin{minipage}{0.48\textwidth}
\centering
\vspace{-3pt}
\includegraphics[width=0.92\textwidth]{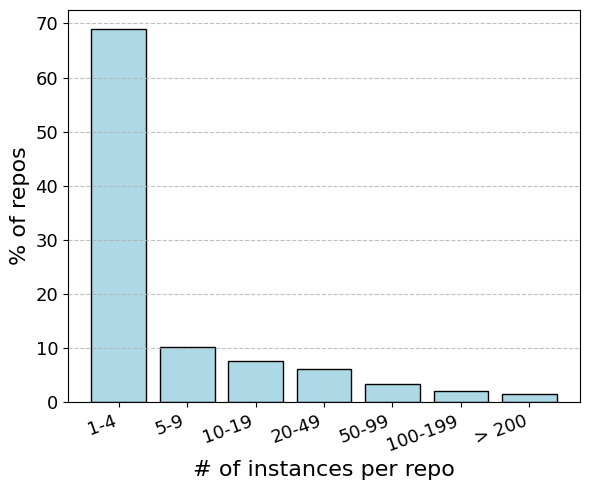}
\vspace{-5pt}
\label{fig:data-distribution}
\end{minipage}
\caption{Dataset statistics for our training dataset (left) and histogram of repo-level instance frequencies (right). Except for size, we report the instance-level averages for all metrics in the table.}
\label{table:dataset-stats-combined}
\vspace{-2mm}
\end{table*}

\begin{figure*}[htbp]
\centering
\begin{subfigure}{0.48\textwidth}
    \centering
    \includegraphics[width=\textwidth]{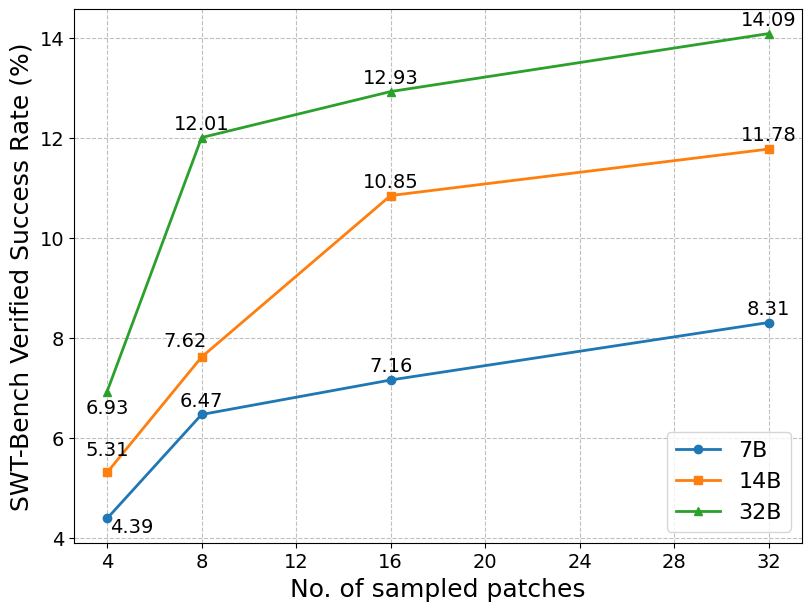}
    \caption{Success Rate ($\mathcal{S}$)}
\end{subfigure}
\hfill
\begin{subfigure}{0.48\textwidth}
    \centering
    \includegraphics[width=\textwidth]{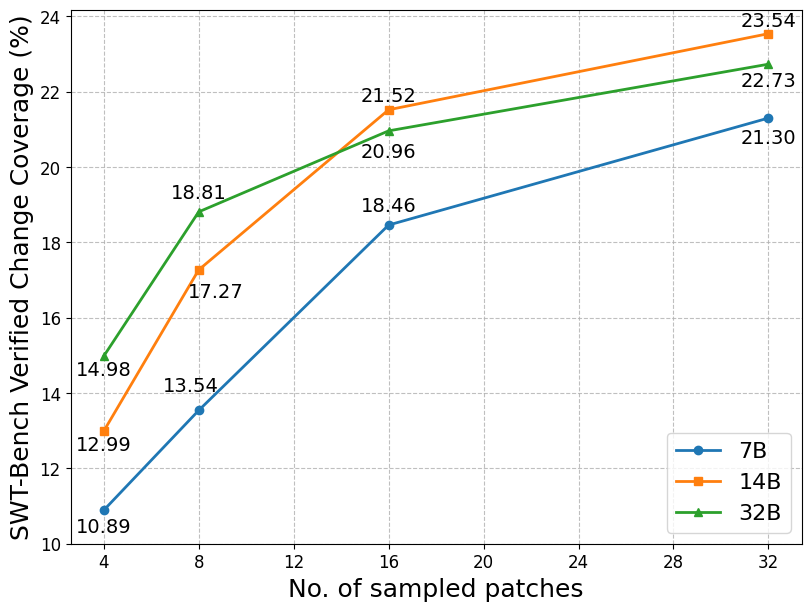}
    \caption{Change Coverage ($\Delta\mathcal{C}$)}
\end{subfigure}
\caption{Inference-time scaling on SWT-Bench Verified. Increasing sampled candidate patches consistently improves performance without saturation.}
\label{fig:inference_scaling}
\end{figure*}

\appendix

\section{Appendix}

\subsection{Prompt Template for Issue Reproduction}\label{app:prompt_issue_reproduction}

We provide the prompt used for both training and inference. Training uses gold source and test files, while inference uses outputs from the localization pipeline (\S\ref{sec:code_localization}).

\begin{tcolorbox}[
  title=Issue Reproduction Prompt,
  colback=blue!5!white,
  colframe=black,
  fonttitle=\bfseries,
  listing only,
  listing options={breaklines=true,basicstyle=\ttfamily\scriptsize}
]
\textbf{System Prompt:}\\
Generate fail-to-pass test edits for a GitHub issue. Do not modify source code. Output must be enclosed in a $<$solution$>$ block and follow SEARCH/REPLACE format.

\textbf{User Prompt:}\\
Given a GitHub issue, source files, and a test file, modify the test file to reproduce the issue.

\textbf{Output Format:}
\begin{itemize}
\item File path
\item $<<<<<<<$ SEARCH
\item Original code
\item =======
\item Modified code
\item $>>>>>>>$ REPLACE
\end{itemize}
\end{tcolorbox}

\subsection{Test File Reranking Prompt}\label{app:prompt_reranking}

We use Qwen3-Embedding \citep{zhang2025qwen3} to rerank BM25-retrieved test files.

\begin{tcolorbox}[
  title=Reranking Prompt,
  colback=blue!5!white,
  colframe=black,
  fonttitle=\bfseries,
  listing only,
  listing options={breaklines=true,basicstyle=\ttfamily\scriptsize}
]
\textbf{Instruction:} Given a GitHub issue and a test file, determine whether it contains sufficient context (classes, functions, boilerplate) to support adding fail-to-pass tests that reproduce the issue.

\textbf{Query:} \{Issue description\}
\end{tcolorbox}

\subsection{Dataset Filtering Rationale}\label{app:filtering_rationale}

We apply filtering to improve data quality under compute constraints:

\begin{enumerate}[leftmargin=*]
\item Remove SWT-Bench repositories to avoid leakage.
\item Exclude non-Python edits (e.g., docs, configs).
\item Limit to $\le 3$ source files to reduce complexity.
\item Keep PRs with exactly one test file, following prior work \citep{xie2025swefixer}.
\item Remove empty issue descriptions due to missing supervision signal.
\end{enumerate}

\subsection{Training Details}\label{app:training_details}

We train for 2 epochs using batch size 32, AdamW \citep{loshchilov2019decoupledweightdecayregularization}, cosine LR scheduling, and gradient clipping (norm 1.0). Learning rates are $1e^{-5}$ for Gemma-3-12B and $3e^{-5}$ for others. Training uses \texttt{torchtune} or \texttt{Llama-Factory} (Gemma). Experiments run on 2–8× A100 GPUs.

\subsection{Reranking Strategy}\label{app:reranking_options}

We observe that change coverage ($\Delta\mathcal{C}$) is 2–3× higher for successful patches than failed ones, making it a stronger ranking signal than self-consistency. In practice, this can be estimated using candidate source patches from systems like Agentless \citep{xia2024agentless}.

\subsection{Oracle Retrieval Evaluation}\label{app:oracle_results}

We evaluate edit-only capability using oracle retrieval (gold source and test files provided). We sample with temperature 0 and retry up to 10 times with incremental temperature increases.
\begin{table}[ht]
\centering
\small
\resizebox{\columnwidth}{!}{
\begin{tabular}{l|cc|cc|cc}
\toprule
\textbf{Model} & \multicolumn{2}{c|}{\textbf{Applicability}} & \multicolumn{2}{c|}{\textbf{Coverage}} & \multicolumn{2}{c}{\textbf{Success}} \\
 & base & SFT & base & SFT & base & SFT \\
\midrule
Qwen2.5-7B  & 40.42 & 84.99 & 3.23 & 21.30 & 1.15 & 8.31 \\
Qwen2.5-14B & 30.95 & 84.99 & 2.51 & 23.54 & 1.39 & 11.79 \\
Qwen2.5-32B & 84.76 & 84.76 & 19.14 & 22.73 & 9.93 & 14.09 \\
\bottomrule
\end{tabular}
}
\caption{Oracle retrieval evaluation on SWT-Bench Verified.}
\label{tab:oracle_performance}
\end{table}

\subsection{Training Data Scaling}\label{app:train_scaling}

We train on 5\%, 25\%, 50\%, and 100\% of 23K instances under oracle retrieval.

\begin{figure*}[htbp]
\centering
\begin{subfigure}{0.48\textwidth}
    \centering
    \includegraphics[width=\textwidth]{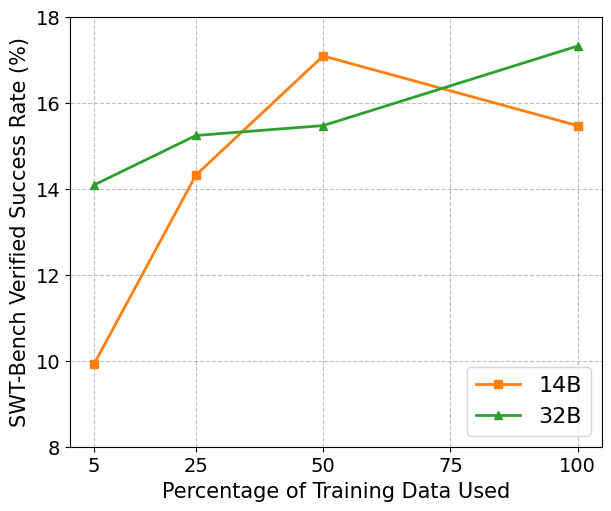}
    \caption{Success Rate}
\end{subfigure}
\hfill
\begin{subfigure}{0.48\textwidth}
    \centering
    \includegraphics[width=\textwidth]{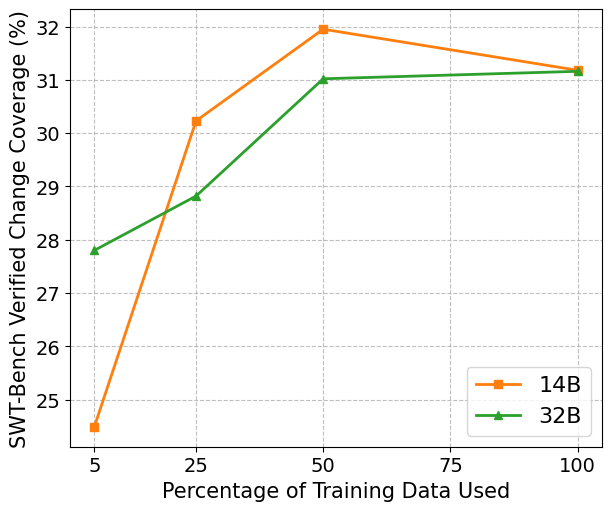}
    \caption{Change Coverage}
\end{subfigure}
\caption{Scaling trends with increasing training data.}
\label{fig:train_scaling}
\end{figure*}

Performance improves consistently with data size. The 14B model saturates around 50\%, while 32B continues improving, showing higher capacity. Even 5\% data yields strong gains.

\subsection{Prior Benchmark Comparison}\label{app:prior_methods}

We compare against closed-source and open models on SWT-Bench Verified.

\begin{table}[ht]
\centering
\small
\begin{tabular}{lcc}
\toprule
\textbf{Method} & \textbf{Model} & \textbf{Success} \\
\midrule
ZeroShotPlus & GPT-4o & 14.30 \\
LIBRO        & GPT-4o & 17.80 \\
\midrule
SWE-Tester 7B  & Qwen2.5-7B  & 8.31 \\
SWE-Tester 14B & Qwen2.5-14B & 11.79 \\
SWE-Tester 32B & Qwen2.5-32B & \textbf{14.09} \\
\bottomrule
\end{tabular}
\caption{Comparison on SWT-Bench Verified.}
\label{tab:baseline_results}
\end{table}

Our 32B model is competitive with GPT-4o-based pipelines despite being fully open-weight.

\subsection{PEFT Evaluation}\label{app:peft_eval}

We evaluate LoRA (r=8, $\alpha=16$) on Qwen2.5-Coder models.

\begin{table}[ht]
\centering
\small
\resizebox{\columnwidth}{!}{
\begin{tabular}{l|cc|cc|cc}
\toprule
Model & \multicolumn{2}{c|}{Applicability} & \multicolumn{2}{c|}{Coverage} & \multicolumn{2}{c}{Success} \\
 & base & SFT & base & SFT & base & SFT \\
\midrule
LoRA-7B  & 10.62 & 69.28 & 2.03 & 25.74 & 0.69 & 10.39 \\
LoRA-32B & 73.74 & 68.12 & 24.55 & 29.16 & 11.32 & 16.86 \\
\bottomrule
\end{tabular}
}
\caption{LoRA fine-tuning under oracle retrieval.}
\label{tab:lora_performance}
\end{table}
PEFT yields strong gains, showing full fine-tuning is not required.

\subsection{Repository Breakdown}\label{app:repo_breakdown}

We report per-repository success rates for Qwen2.5-Coder models.

\begin{table}[ht]
\centering
\small
\begin{tabular}{lccc}
\toprule
Repository & 7B & 14B & 32B \\
\midrule
astropy & 5.88 & 23.53 & 23.53 \\
django & 7.87 & 10.19 & 11.11 \\
matplotlib & 0.00 & 9.38 & 18.75 \\
seaborn & 0.00 & 0.00 & 0.00 \\
flask & 0.00 & 100.00 & 100.00 \\
requests & 25.00 & 0.00 & 0.00 \\
xarray & 26.67 & 6.67 & 6.67 \\
pylint & 0.00 & 16.67 & 33.33 \\
pytest & 13.33 & 13.33 & 26.67 \\
scikit-learn & 16.67 & 25.00 & 25.00 \\
sphinx & 3.57 & 3.57 & 3.57 \\
sympy & 8.22 & 13.70 & 16.44 \\
\bottomrule
\end{tabular}
\caption{Repository-wise performance breakdown.}
\label{tab:model-repo-percentages}
\end{table}

\subsection{Dataset License}

All data is derived from publicly available GitHub repositories under permissive licenses (MIT/Apache/BSD). We respect original licensing constraints and release curation scripts for compliance.
\end{document}